\documentclass[%
 aip,
 amsmath,amssymb,amsthm,
preprint,%
]{revtex4-1}

\usepackage{graphicx,float}
\usepackage{dcolumn}
\usepackage{bm}
\usepackage{color}

\usepackage[utf8]{inputenc}
\usepackage[T1]{fontenc}
\usepackage{mathptmx}

\usepackage[notref,notcite]{showkeys}

\newtheorem{thm}{Theorem}[section]

\newtheorem{lem}[thm]{Lemma}
\newtheorem{cor}[thm]{Corollary}

\newtheorem{proof}[thm]{Proof}

\newcommand{\ie}{{\it i.e.}}
\newcommand{\eg}{{\it e.g.}}

\begin{document}


\title[Exponential Integrators]{Exponential Integrators for Stochastic Schr\"{o}dinger Equation } 

%

\author{Jingze Li}
\email{201611130119@mail.bnu.edu.cn}
\affiliation{School of Mathematical Sciences,\\  Beijing Normal University, Beijing 100875, P.R.China.
}
\author{Xiantao Li}
\email{xiantao.li@psu.edu}
\affiliation{Department of Mathematics,\\ The Pennsylvania State University, University Park, PA 16802, USA.
}%


%


\begin{abstract}
We present a class of exponential integrators to compute solutions of the stochastic Schr\"{o}dinger equation arising from the modeling of 
open quantum systems. In order to be able to implement the methods within the same framework as the deterministic 
counterpart, we express the solution using the Kunita's representation. With appropriate truncations, the solution operator can be written as matrix exponentials, which can be efficiently implemented by 
the Krylov subspace projection. The accuracy is examined in terms of the strong convergence, by comparing trajectories, and the weak convergence, by comparing  the density-matrix operator.  We show that the local accuracy can be further improved by introducing a third-order commutator in the exponential.  The effectiveness of the proposed methods is tested using the example from Di Ventra et al. [Journal of Physics: Condensed Matter, 2004]. 

 
\end{abstract}

\maketitle

%

\section{Introduction}

The modeling of open quantum systems has been a subject of immense interest for decades \cite{breuer2002theory,carmichael2009open,weiss2012quantum}. The primary focus is  on quantum systems coupled to the environment.  
While direct computation based on the entire system is infeasible,  reduced models where the influence from the bath is implicitly incorporated have shown great promises. One remarkable approach is the stochastic Schr\"{o}dinger equation (SSE), which can be formally derived from the Schr\"{o}dinger equation for the entire system by using a projection formalism \cite{gaspard1999non}, together with a Markovian approximation. On the other hand, the dynamics of the density-matrix follows a Liouville von Neumann equation that agrees with the Lindblad equation \cite{lindblad1976generators}. Therefore, it can be used as a computational approach to obtain the density-matrix, especially when the dimension of the problem is high.   The SSE has been used in quantum transport to study non-equilibrium  transport problems by Di Ventra's and coworkers \cite{di2004transport} as well as the extension to time-dependent density-functional theory  \cite{di2007stochastic,d2008stochastic}. The  recent review  \cite{biele2012stochastic} surveyed many of these aspects.

This paper is primarily concerned with the numerical treatment of the SSE. In the deterministic case, \ie, the time-dependent Schr\"{o}dinger equation (TDSE), many numerical methods are available. Typically, due to the large number of degrees of freedom in many practical applications, the efficiency has been an important focus in selecting an appropriate method. In addition, the time reversibility and the unitary property of the evolution operator, are also desired. These considerations seem to deem classical Runge-Kutta methods unfavorable. Finally, since the time scale associated with electron dynamics is often on the order of attoseconds, there is a stringent limit on the size of the time steps. Although many implicit methods can greatly mitigate this issue, the implementation is often not straightforward.  By a comparative study of some existing methods,  Castro el al. \cite{castro2004propagators} demonstrated that numerical methods can be constructed based on the exponential representation of the solution operator. The Krylov subspace method with Lanczos orthogonalization has been the  most efficient in terms of the overall computation cost \cite{castro2004propagators}. 
This technique approximates the matrix exponential by projecting it onto a subspace \cite{saad1992analysis}, and the problem is reduced to computing the matrix exponential of a smaller matrix which can be easily tackled by many existing methods \cite{gene1996matrix}. 

Unlike the deterministic case, not many methods have been developed particularly for the SSE. Some of the subtleties in treating
stochastic models numerically have been explained in \cite{burrage2006comment}. The classical Euler-Maruyama and Milstein methods \cite{kloeden2013numerical} are simple extensions of the Runge-Kutta methods in solving stochastic models, and similar to the deterministic case, they may not be well suited for SSEs. Many of the standard higher-order methods \cite{kloeden2013numerical} are quite involved in the case of system of equations with multiplicative noise.  In this paper we propose to extend 
the framework of exponential integrators  for deterministic systems \cite{hochbruck1997krylov,castro2004propagators} to the SSE. We first express the solution operator using the Kunita's notation \cite{kunita1980representation} in the context of stochastic differential equations (SDEs), where the operator in the exponential consists of an infinite series of  commutators and multiple stochastic integrals. With truncations, we obtain approximations of the solution operator. We then show that once the stochastic noise is realized, the truncated operator represents deterministic Schr\"{o}dinger equations, and a matrix exponential can be used to represent the solution. At this point, the Krylov subspace method can again be used.  Since the Hamiltonian in this case is no longer Hermitian, we will use the Arnodi's algorithm \cite{saad2003iterative} to obtain the orthogonal basis.   

We also studied the order of accuracy of the proposed methods. We follow two tracks: the strong convergence, where the approximate solution is compared to the true solution on a 
trajectory-wise basis, and the weak convergence, where we study the accuracy in terms of the density-matrix. This is particularly important since physical observables, \eg, electron density and current, can be directly obtained from the density matrix.  
 
The remaining part of this paper is organized as follows: in Section II, we present the theoretical results. We start by defining the solution operator for the SSE, and then derive the exponential integrator and the approximation scheme. We examine the accuracy and the extension to nonlinear problems. In Section III ,we present the numerical results  to demonstrate the effectiveness of the exponential schemes.

\section{Theory and Methods}

We consider 
 an SSE as follows,
\begin{equation}\label{eq: sse} 
\begin{aligned}
   d\psi(\bold{r},t) &=  (-i\hat{H}-\frac{1}{2}\hat{V}^*\hat{V})\psi(\bold{r},t)dt + \hat{V}\psi(\bold{r},t)dW_t,
    \\
    \psi(\bold{r},0)& = \psi_0.
\end{aligned}
\end{equation}
Here we have chosen to write the SDEs in the conventional form \cite{oksendal2003stochastic}, where solutions are interpreted in integral forms. 
In \eqref{eq: sse},   $\psi(\bold{r},t)$ is the wave function in an appropriate Hilbert space. Typically the system has multiple orbitals, each of which would satisfy an equation of this form; But  it suffices to describe the case with a single wave function.   In the equation \eqref{eq: sse}, $\hat{H}$ is a Hermitian operator for the Hamiltonian,  and $\hat{V}$ is the bath operator.  $W_t$ is the standard one-dimensional Wiener process. Formally $\ell(t)\doteq \frac{dW_t}{dt}$  can be interpreted as a white noise. In applications, the system could be coupled with multiple environments; then there would a set of heat baths $\hat{V}_{\alpha}$, corresponding to a set of stochastic noises $\{l_{\alpha}(t)\}$. The Markovian assumption embodies the following properties,
\begin{equation} 
\begin{aligned}
\overline{\ell_{\alpha}(t)} &=0,\\
\overline{\ell_{\alpha}(t)\ell_{\beta}(t')} &= \delta_{\alpha,\beta}\delta(t-t'),
\end{aligned}
\end{equation}
where the over line indicates the stochastic average over an ensemble of realizations of the Brownian motion. Finally,  as emphasized 
numerous times in the literature, the SSE \eqref{eq: sse} is interpreted in the It\^{o} sense \cite{oksendal2003stochastic}. 

 In the following discussion, we generally assume that the Hamiltonian $\hat{H}$ is linear. In the case where $\hat{H}$ is nonlinear, 
 we will adopt the operator-splitting method (\eg see \cite{misawa2001lie,watanabe2002efficient}), which separates the Hamiltonian into linear and nonlinear parts. This will be discussed in details in Section II.C. 
 Throughout this paper, we also assume that the operators $\hat{H}$ and $\hat{V}$ are discretized spatially.

Before we present approximation schemes, we first discuss how the {\it exact} solution can be represented.

\subsection{Solution operators for general deterministic and stochastic systems}

We start by considering the deterministic case, \ie, where the stochastic bath operator $V\equiv 0$. Inspired by the idea of the Koopman operator\cite{koopman1931hamiltonian,brunton2019data,mauroy2016linear}, we give a representation of the solution in the form of an exponential operator.  More specifically, for a general $n-$dimensional autonomous dynamical system, 
\begin{equation}
\dot{\bold{x}} = \bold{F}(\bold{x}), \quad \bold{x}(0)= \bold{x}_0,
\end{equation}
the Koopman operator $\mathcal{U}(t)$ describes the evolution of an observable $A$,
\[  A(\bold{x}(t)) = \mathcal{U}(t) A(\bold{x}_0).\]
It can be expressed in an exponential form,
\begin{equation}
 \mathcal{U}(t) = e^{\mathcal{L}t},  \quad \mathcal{L} = \bold{F} (\bold{x}_0)\cdot \nabla_{\bold{x}_0}.
\end{equation}

By applying this result to the deterministic Schr\" {o}dinger equation, we have the following formula.

\begin{lem}

When  $\hat{V}\equiv 0$, the solution of  the SSE  \eqref{eq: sse} can be represented by  the exponential operator,
\begin{equation}\label{eq: koopman}
  \psi(\bold{r},t) = \exp(\hat{D}_t)\psi_0,
\end{equation}
where $\hat{D}_t$ is a differential operator, given by,
\begin{equation}\begin{aligned} 
   \hat{D}_t =&&-it\hat{H}(\psi_0\frac{\partial }{\partial \psi_0} - \psi^*_0\frac{\partial }{\partial \psi^*_0} ).
\end{aligned}\end{equation}
Here $\psi^*$ denote the complex conjugate of $\psi$.

Furthermore, due to the linearity of the SSE  \eqref{eq: sse}, the solution can also be expressed as a matrix/operator exponential,
\begin{equation}\begin{aligned} 
   \psi(\bold{r},t) = && \exp( -it\hat{H})\psi_0.
\end{aligned}\end{equation}
The matrix exponential is defined by the Taylor series of the exponential function. 
\end{lem}

A short derivation can be found in Appendix A. We should point out that  in principle, the Koopman's solution form \eqref{eq: koopman} holds for general  nonlinear systems, and it can be regarded as the foundation of operator-splitting algorithms, \eg, those for classical molecular dynamics models\cite{watanabe2002efficient}.
Although it is not directly relevant to the discussions here, the Koopman operator can be applied to functions of the state variable. For non-autonomous systems, the Koopman
operator can be extended by including the term $\partial_{t_0}$, where $t_0$ is the initial time.

Now we turn to the stochastic case. We first introduce the Kunita's results for a  general SDE  \cite{kunita1980representation}.  More specifically, we have:
\begin{thm}[Kunita 1980, Lemma 2.1]
For  an autonomous SDE of Stratonovich type,
\begin{equation}
dz_t = a(z_t)dt + b(z_t)\circ dW_t,
\end{equation}
the solution with an initial value $z(0) = z_0$ can be represented as,
\begin{equation}\begin{aligned} 
  z(t) = && \exp(\hat{D}_t)z_0,
\end{aligned}\end{equation}
where the differential operator $\hat{D}_t$ is given by 
\begin{equation}\begin{aligned} 
  \hat{D}_t = && t X_0 + W_t X_1 +\frac{1}{2}(J_{(0,1)} - J_{(1,0)})[X_0,X_1] + \sum_{J;3\leq |J|}\{\sum_{\Delta J}^* c_{\Delta J}W^{\Delta J}(t)\}X^{J}.
\end{aligned}\end{equation}
\end{thm}

Here $X_0$ and $X_1$ are differential operators defined by
\begin{equation}\begin{aligned} 
  X_0 \doteq \sum_{i=1}^n a^i\partial_i , \quad X_1\doteq  \sum_{i=1}^n b^i\partial_i, \quad \partial_i \doteq \frac{\partial}{\partial z_{0,i}},
\end{aligned}\end{equation}
In addition, $J_{(0,1)}$ and $J_{(1,0)}$ are Stratonovich stochastic integrals defined respectively as
\begin{equation}\begin{aligned} 
  J_{(0,1)} &\doteq  \int_0^t sdW_s,\\
  J_{(1,0)} &\doteq  \int_0^t W_sds.
\end{aligned}\end{equation}

The notation $[X_0,X_1]$ is the usual Lie bracket defined by $X_0X_1 - X_1X_0$;  $J = (j_1,\ldots,j_m)$ indicates multi-indices and $X^{J} = [\cdots[X_{j_1},X_{j_2}]\cdots X_{j_m}]$ are
high-order commutators. The rest of the notations have been defined explicitly in Kunita's work \cite{kunita1980representation}. 

Note that in this lemma the stochastic integrals are of Stratonovich type\cite{oksendal2003stochastic}.  But SDEs of this type can be converted from (and to)  It\^{o} SDEs. More specifically, for a general $n$-dimensional It\^{o} SDEs
\begin{equation}\begin{aligned} 
  dz_{t} = && a(t,z_t)dt + b(t,z_t)dW_t,
  \end{aligned}\end{equation}
where $a,z_t\in \mathbb{R}^n$,$b\in \mathbb{R}^{n\times m}$ and $W_t\in \mathbb{R}^m$. The corresponding Stratonovich SDE is given by\cite{oksendal2003stochastic}
\begin{equation}\begin{aligned} 
  dz_{t} = && \underline{a}(t,z_t)dt + b(t,z_t)\circ dW_t.
  \end{aligned}\end{equation}
Here the modified drift term is defined componentwise by
\begin{equation}\begin{aligned} 
\underline{a}^i(t,z) = && a^i(t,z) - \frac{1}{2}\sum_{j=1}^n\sum_{k=1}^mb^{j,k}(t,z)\frac{\partial b^{i,k}}{\partial x_j}(t,z).
  \end{aligned}\end{equation}

Therefore, in order to use the Kunita's notation,   we first need to switch our It\^{o} type SSE \eqref{eq: sse} to the corresponding Stratonovich type,
\begin{equation}\begin{aligned} \label{eq: sse'} 
 d\psi(\bold{r},t) =&&  \big(-i\hat{H}-\frac{1}{2}(\hat{V}^*+ \hat{V})\hat{V}\big)\psi(\bold{r},t)dt + \hat{V}\psi(\bold{r},t)\circ dW_t.
\end{aligned}\end{equation}
Effectively, this introduces the additional drift term $-\frac12\hat{V}^2 \psi$.

  Now we apply the Kunita's lemma\cite{kunita1980representation} to  \eqref{eq: sse'}, and with direct computation we have,    
\begin{cor}
The solution of  the equation \eqref{eq: sse} (or equivalently \eqref{eq: sse'}) 
can be represented as,
\begin{equation}\begin{aligned} 
   \psi(\bold{r},t) = && \exp(\hat{D}_t)\psi(\bold{r},0),
\end{aligned}\end{equation}
where the solution operator $\hat{D}_t$ can be approximated by,  
\begin{equation} 
  \hat{D}_t \approx \hat{D}^I_t \doteq \big (-i\hat{H}- \frac{1}{2}(\hat{V}^*+ \hat{V} )\hat{V} \big)t + \hat{V}W_t,
\end{equation}
by  keeping the first-order Stratonovich integrals,  and
\begin{equation}
  \hat{D}_t \approx \hat{D}^{II}_t 
  \doteq \big( -i\hat{H}- \frac{1}{2}(\hat{V}^*+ \hat{V} )\hat{V}\big )t + \hat{V}W_t+  \frac{1}{2}\big(J_{(0,1)}(t)-J_{(1,0)}(t) \big) (\frac{1}{2}[\hat{V}^*, \hat{V}] \hat{V}+ i[\hat{H},\hat{V}]).
\end{equation}
  by retaining the second-order Stratonovich integrals. 
\end{cor}
Interested readers can find the detailed derivation in the Appendix B.

\subsection{Exponential integrators}
Based on the truncated Kunita's notation of solution operator, one can construct numerical methods \cite{misawa2000numerical,misawa2001lie,telatovich2017strong}. Here we  focus on one-step methods, where the solution at the next step is updated  only based on the solution at the current step \cite{kloeden2013numerical}. The same procedure would be repeated at each time step. In this case, it is enough to illustrate the methods within one step, \eg, from $t=0$ to $t=\Delta t,$ with $\Delta t$ 
being the step length. To this end, we first sample $W_t$ and denote,
\begin{equation}
 \Delta W= W_{\Delta t} - W_0. 
\end{equation}
For later steps, we define $  \Delta W_n= W_{t_{n+1}} - W_{t_n},$ and simply replace $\Delta W$ by $\Delta W_n$ and apply the same procedure. In the following context we also use $\psi(t_n) \approx \psi_n$ as the numerical approximation of the state vector $\psi$ at time $t_n$.

With this notation, the first-order truncation, when applied to the initial condition, becomes,
 \begin{equation}\begin{aligned} 
    \hat{D}^I_{\Delta t} &= \big((-i\hat{H}- \frac{1}{2}(\hat{V}^*+ \hat{V} )\hat{V} )\Delta t + \hat{V}\Delta W \big)\psi\frac{\partial }{\partial \psi}\\
   &+\big ( (i\hat{H}- \frac{1}{2}(\hat{V}^*+ \hat{V} )\hat{V} )\Delta t + \hat{V}\Delta W \big)\psi^*\frac{\partial }{\partial \psi^*}.
 \end{aligned}\end{equation} 
 
Since  $\Delta W$ has been realized, the operator above can be viewed as a deterministic operator, and in light of the Koopman's notation \eqref{eq: koopman}, it generates a solution of the following ODEs, 
\begin{equation}
  \frac{\partial}{\partial \tau} \phi =  \big(-i\hat{H}- \frac{1}{2}(\hat{V}^*+ \hat{V} )\hat{V}\big )\phi\Delta t + \hat{V}\phi\Delta W, \quad \tau \in [0,\Delta t].
  \end{equation}

In particular,  Lemma II.1 implies that the solution can be written as an exponential,
 \begin{equation} \label{eq: exp1} 
  \phi(\Delta t) = \exp(-i\Delta t \hat{\mathcal{H}}_{I})\phi(0),
 \end{equation}
 where
 \begin{equation}
 \hat{\mathcal{H}}_{I} =  \hat{H}- \frac{i}{2}(\hat{V}^*+ \hat{V} )\hat{V}  + i\frac{\Delta W}{\Delta t}\hat{V}.
  \end{equation}
  
Using \eqref{eq: exp1} we construct the follow exponential scheme,
\begin{equation}
 \psi_{n+1}=  \exp(-i\Delta t \hat{\mathcal{H}}_{I}) \psi_n.
\end{equation}
 We will later refer to this Scheme as \textbf{Scheme I}. The matrix exponential in \eqref{eq: exp1} will be treated by using Krylov subspace projection method together with the Arnoldi's method\cite{saad2003iterative,hochbruck1997krylov}. In general, this algorithm yields,

\begin{equation}\begin{aligned} 
   A &\approx  V_mH_mV_m^*,
\end{aligned}\end{equation}
where $m$ is the dimension of the Krylov subspace, $H_m\in \mathbb{C}^{m\times m}$ is a Hessenberg matrix, and $V_m\in \mathbb{C}^{N\times m}$ consists of $m$ orthonormal column vectors. Thus the matrix exponential is approximated by
\begin{equation}\begin{aligned} 
   \exp(A)v &\approx  V_m\exp(H_m)e_1,
\end{aligned}\end{equation}
where $e_1$ is the first unit vector in $\mathbb{C}^m$. 

Since $m$ is relatively small, $\exp(H_m)$ is much easier to compute than the original matrix exponential and can be computed by any of the current methods that computes a matrix exponential \cite{gene1996matrix}. 
In our numerical experiment (Section III),  even $m = 3 $  is adequate. 
%
  
  We should note that although in  Scheme I,  we only take the first order terms in the exponent, this  method is different from the Euler-Maruyama method. In fact, a direct expansion 
  yields the term 
  $\frac{1}{2}(\nabla b)b\Delta W^2$,
  which appears in the Milsteins scheme, a first order method (rather than 0.5 order). 
  
  \medskip
  
Similarly, we construct an integrator  by the truncation of the solution operator $\hat{D}_t$ up to the second order Stratonovich integral,
 \begin{equation}\begin{aligned} 
       \hat{D}^{II}_{\Delta t} &= [ \{-i\hat{H}- \frac{1}{2}(\hat{V}^*+ \hat{V} )\hat{V}\}\Delta t + \hat{V}\Delta W+  \Delta U (\frac{1}{2}[\hat{V}^*, \hat{V}] \hat{V}+ i[\hat{H},\hat{V}])]\psi\frac{\partial }{\partial \psi}\\
   & +[ \{i\hat{H}- \frac{1}{2}(\hat{V}^*+ \hat{V} )\hat{V}\}\Delta t + \hat{V}\Delta W+  \Delta U (\frac{1}{2}[\hat{V}^*, \hat{V}] \hat{V}- i[\hat{H},\hat{V}])]\psi^*\frac{\partial }{\partial \psi^*},
 \end{aligned}\end{equation}
 where $\Delta U = \frac{1}{2}(J_{(0,1)}-J_{(1,0)}) $ is a Gaussian random variable with mean  0 and 
variance $\frac{\Delta t^3}{12}$.
 
 Once $\Delta W$ and $\Delta U$ are realized, the solution corresponds to that of the following ODEs 
 \begin{equation}\begin{aligned} 
  \frac{\partial}{\partial \tau}   \psi & = \Big (-i\hat{H}- \frac{1}{2}(\hat{V}^*+ \hat{V} )\hat{V}\Big )\psi\Delta t + \hat{V}\psi\Delta W+  (\frac{1}{2}[\hat{V}^*, \hat{V}] \hat{V}+ i[\hat{H},\hat{V}])\psi \Delta U, \quad \tau \in [0,\Delta t].
 \end{aligned}\end{equation}
 
 An exponential scheme can then be constructed accordingly:
 \begin{equation} 
  \psi_{n+1} = \exp(-i\Delta t \hat{\mathcal{H}}_{II})\psi_n, 
 \end{equation}
 where the matrix  $\hat{\mathcal{H}}_{II}$ is given by
  \begin{equation} 
 \hat{\mathcal{H}}_{II} =  \hat{H}- \frac{i}{2}(\hat{V}^*+ \hat{V} )\hat{V}  + i\frac{\Delta W}{\Delta t}\hat{V}+  i\frac{\Delta U}{\Delta t} (\frac{1}{2}[\hat{V}^*, \hat{V}] \hat{V}+ i[\hat{H},\hat{V}]).
 \end{equation}
 We will later refer to this Scheme as \textbf{Scheme II}. Scheme II has one more  term in the exponential  than Scheme I.
 
The higher order Stratonovich integral terms from the Kunita's expansion is complicated. But we discovered that by incorporating two more commutator terms in our truncation, we get better convergence results with respect to the density-matrix operator. This will be referred to as \textbf{Scheme III}. It is as follows,
 \begin{equation} 
  \psi_{n+1}= \exp(-i\Delta t \hat{\mathcal{H}}_{III})\psi_n, 
 \end{equation}
 where 
  \begin{equation}\begin{aligned} 
 \hat{\mathcal{H}}_{III} =&\hat{\mathcal{H}}_{II} 
 +  i\Delta t \Big(\frac{1}{24}[\hat{V},[\hat{V}^*,\hat{V}]\hat{V}]+\frac{i}{12}[\hat{V},[\hat{H},\hat{V}]]\Big).
 \end{aligned}\end{equation}

 \subsection{The Extension to Nonlinear SSEs}
 
 In the discussion above we have assumed that the Hamiltonian is linear and independent of time. The extension to nonlinear problems, \eg, those that resemble the  Kohn-Sham equations in the time-dependent density-functional theory \cite{marques2006time,runge1984density} with an external potential, is straightforward.

 Following Watanabe and Ksukada \cite{watanabe2002efficient}, we can separate the Hamiltonian as 
\begin{equation}
H = H_{0} + H_{1}(t),
\end{equation}
where $H_{0}$ is the linear part and $H_{1}(t)$ contains the nonlinear contribution. We assume that the nonlinearity appears in the potential as a local operator. 
Then, a one-step can method be constructed using an operator-splitting method \cite{watanabe2002efficient}, 
\begin{equation}
\psi_{n+1} =\exp(\frac{\Delta t}{2} \partial_{t_n}) \exp(-i\frac{\Delta t}{2}H_1)\exp(-i\Delta t H_{0})\exp(-i\frac{\Delta t}{2}H_1) \exp(\frac{\Delta t}{2} \partial_{t_n}) \psi_n.
\end{equation}
The operator $\partial_{t_n}$ operates on quantities that explicitly depend on the time variable. 
The exponential associated with the linear Hamiltonian has been discussed in the previous section. On the other hand, the exponential for the nonlinear part, due to the fact that $H_1$ is diagonal, is also straightforward. 
The error associated with the splitting, which can be analyzed using the Baker-Campbell-Hausdorff (BCH) formula \cite{yoshida1990construction},  is locally of the order $\mathcal{O}(\Delta t^3).$

\subsection{The accuracy of the exponential integrators}

 Now we discuss the accuracy of our schemes as $\Delta t\rightarrow 0$. Unlike the deterministic case, the convergence of numerical methods for stochastic models can be formulated in both the strong and weak sense \cite{kloeden2013numerical}.
 
 \noindent \textbf{Strong convergence.}   
The strong convergence of these schemes is summarized as follows:
\begin{thm}
Let $\psi_T$ be the exact solution of model (1) at time $T$. Let $\tilde{\psi}_T^{\Delta t}$ be the approximation by the exponential integrator discussed above at time $T$, with time discretization $\Delta t$. 
Then
\begin{equation}
\overline{ |\tilde{\psi}_T^{\Delta t} - \psi_T| } \leq  K_1 \Delta t^{\gamma}
\end{equation}
holds, where the constants $K_1, K_2$ do not depend on $\Delta t$. Here $\gamma = 1$ for Scheme I, and $\gamma = 1.5$ for Scheme II.
 \end{thm}

Following the idea in the proof of Theorem 2.1 in \cite{lord2008efficient},
we can verify the accuracy by comparing the schemes to the stochastic Taylor expansion and utilizing Lemma 5.7.3 and Theorem 10.6.3 in \cite{kloeden2013numerical}.

\
~\\
\noindent\textbf{Weak Convergence.} Following the notations in Kleoden and Platern \cite{kloeden2013numerical}, the weak convergence is
in the sense of averages.  As we have alluded to in the introduction, a primary quantity of interest in open quantum systems is the density-matrix. In the stochastic case, the density-matrix is defined as the following ensemble average \cite{d2008stochastic}, 
\begin{equation} 
\hat{\rho}(t) \doteq \overline{|\psi(t) \rangle \langle\psi(t) | }.
\end{equation}

In general, the analysis of the weak convergence relies on the Dynkin's formula and the backward equation  \cite{kloeden2013numerical}. However, for the SSE \eqref{eq: sse}, one can actually write down an exact equation for the density-matrix, known as the Lindblad equation  \cite{van1992stochastic}

\begin{equation}\label{eq: lindblad}
i \partial_t \rho  =[H,\rho] -\frac{i}{2}(\hat{V}^*\hat{V}\rho + \rho\hat{V}^*\hat{V}-2\hat{V}\rho\hat{V}^*). 
\end{equation}
This can be derived from SSE\eqref{eq: sse} using the It\^{o}'s formula\cite{oksendal2003stochastic} .

With the Lindblad equation, one can expand the density-matrix at time $t=t_n$ as power series of $\Delta t$. 
Meanwhile,  all our schemes can be written in terms of ODEs (with random coefficients) and the corresponding approximate density-matrix 
can also be expanded in the same manner.  With direct comparison, we obtain the order of local consistency error, summarized as follows,

\begin{thm}
If $\rho$ is the exact density matrix and $\hat{\rho}_{I},\hat{\rho}_{II}, \hat{\rho}_{III}$  are the approximations yield by Scheme I, Scheme II and Scheme III, respectively,  then we have
\begin{equation}\begin{aligned} 
  \rho(\Delta t) - \hat{\rho}_{I}(\Delta t) &= \mathcal{O}(\Delta t^2),\\
   \rho(\Delta t) - \hat{\rho}_{II}(\Delta t) &= \mathcal{O}(\Delta t^2),\\
   \rho(\Delta t) - \hat{\rho}_{III}(\Delta t) &= \mathcal{O}(\Delta t^3).
\end{aligned}\end{equation}

\end{thm}
This suggests that Schemes I and II have order 1 weak convergence while Scheme III is second order. Surprisingly, the Scheme II, which has higher strong order, does not have better convergence in terms of the density-matrix. This problem is addressed by 
adding another term to the exponential integrator, and which hence leads to the Scheme III. See Appendix C for details.

\section{Numerical Results}

  

We consider the example used by Di Ventra \textit{et al} \cite{d2008stochastic} to demonstrate the performance of the proposed methods. The 
underlying Schr\"{o}dinger equation,  describing the dynamics of a one-dimensional gas of excited bosons confined in a harmonic potential and in contact with an external bath, is given by,
\begin{equation}\label{eq: SSEC}\begin{aligned} 
  d\psi(x,t) &= -i\Big (-\frac{1}{2m}\frac{d^2}{dx^2}
 +\frac{1}{2}m\omega_0^2x^2+\tilde{g}n(x,t)\Big )\psi(x,t)dt \\
 &-\frac{1}{2}\hat{V}^*\hat{V}\psi(x,t)dt 
  + \hat{V}\psi(x,t)dW_t.
\end{aligned}\end{equation}

Using the same treatment in Di Ventra \textit{et al}\cite{d2008stochastic}, we first pick $\tilde{g} = 0$ and we choose the Hilbert space spanned by the basis set  $\{\varphi_j:j=1\ldots d\}$, consisting of the eigen-functions of the quantum harmonic oscillators. The projection makes the Hamiltonian diagonal, and we choose $d = 20$. To test our schemes, we conducted simulations over the time interval $t\in [0,1]$, with stepsize $\Delta t =  10^{-3}$.
We also take the same bath operator,
\begin{equation}\label{eq: V}
\begin{aligned}
\hat{V} \equiv \delta
 \begin{pmatrix} 
            0 & 1 & 1 &1 & \cdots \\
            0 & 0 & 0 &0 & \cdots\\
            \vdots & \vdots & \vdots &\vdots & \vdots\\
            0 & 0 & 0 & 0 & \cdots
 \end{pmatrix},
\end{aligned}\end{equation}
where $\delta$ is interpreted as a coupling constant.

Before we discuss the convergence results, we should mention that our schemes has shown good numerical stability. We computed the system defined above with different stepsizes, and our schemes is stable for $\Delta t \leq 10^{-3}$, while the Euler-Maruyama method is stable only for $\Delta t \leq 10^{-5}$.

First we examine the strong convergence.  In FIG.2  we compare the following error
 \begin{equation}
 e(t)=  \overline{\| \psi(t)- \hat{\psi}(t)\|},
 \end{equation}
 from the numerical methods.
The expectation is approximated by an average over 100 runs.  Here in our test, the exact solution $\psi$ is computed by the 1.0 order Milstein scheme performed with a  much smaller time step ($\delta t = 10^{-2}\Delta t$).  We observe that  the Scheme II, with expected 1.5 order strong convergence, exhibits a much smaller error than Scheme I, which has 1.0  strong order.

\begin{figure}[H]
\label{fig2}
\centering
 \includegraphics[scale = 0.2]{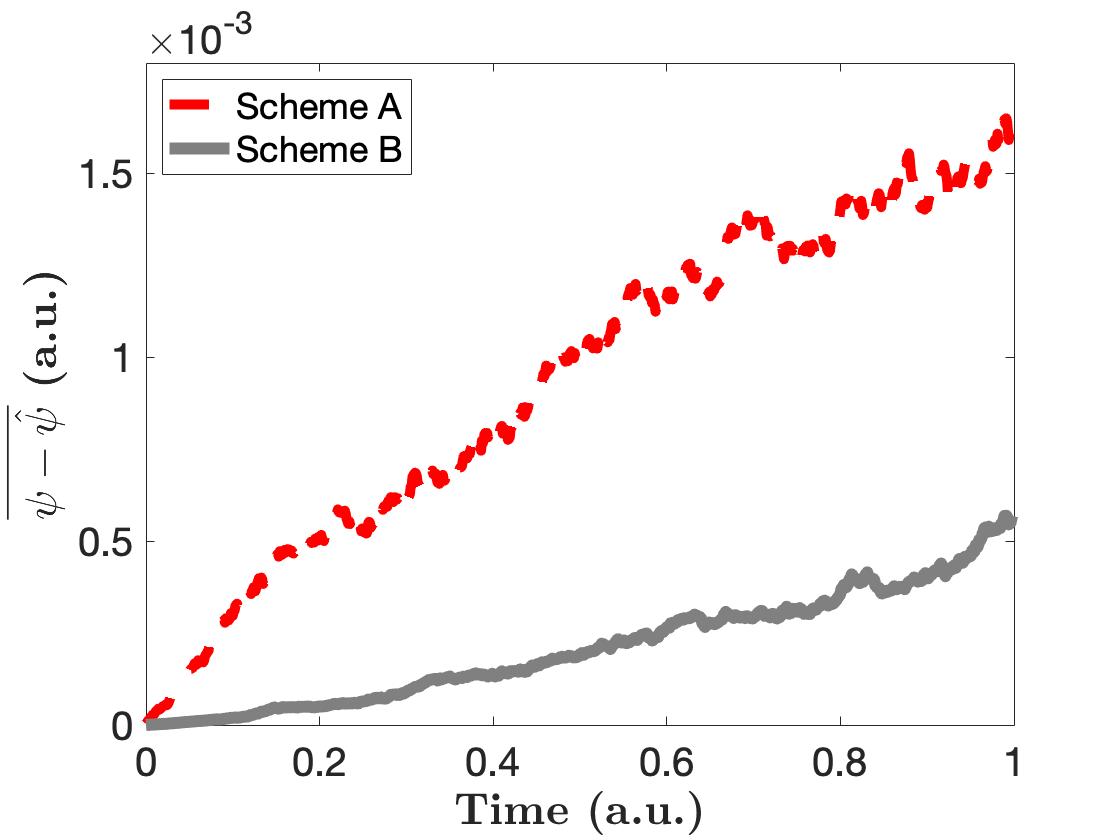}%
 \caption{
 A comparison of the strong order accuracy of schemes I and II. The red dashed line depicts the error from Scheme I, and the gray solid line shows the error from Scheme II. The linear case: $\tilde{g} = 0 $ is considered. } 
 \end{figure}

Next we examine the weak convergence in terms of  the density matrix. In particular, we compare the first entry of the density matrix, which is the square of the coefficient of the ground state when projecting the state vector to the basis of the eigenvectors of the Hamiltonian.   We find that due to the fact that the bath operator $\hat{V}$ \eqref{eq: V} satisfies $[[\hat{V}^*,\hat{V}]\hat{V},\hat{V}] = [[\hat{H},\hat{V}],\hat{V}] = 0$,  Scheme II and Scheme III are identical.  Therefore, we pick another bath operator $\hat{V_1}$, as follows,
 \begin{equation}\label{eq: V1}
 \begin{aligned}
\hat{V}_1 \equiv \delta
 \begin{pmatrix} 
            0 & 1 & 1 &1 & \cdots \\
            0 & 0 & 0 &0 & \cdots\\
            \vdots & \vdots & \vdots &\vdots & \vdots\\
            0& 0 & 0 &0 &0\\
            1 & 1 & \cdots & 1 & 0
 \end{pmatrix}.
\end{aligned}\end{equation}

We consider the error of the first entry of the density matrix $\rho_{11} = \overline{|\psi_0|}^2$ from Scheme I ,  II and III. The results are
displayed in FIG. 3. We approximated the expectation using 100 runs. Again the exact density-matrix $\rho$ is computed by the Milstein scheme with much smaller time step. We observe that Scheme II has a moderate improvement over Scheme I, and Scheme III offers significantly better accuracy.  
This can be attributed to the higher order weak convergence property that we demonstrated in the previous section.

 \begin{figure}[H]
 \centering
 \label{fig3}
 \includegraphics[scale = 0.2]{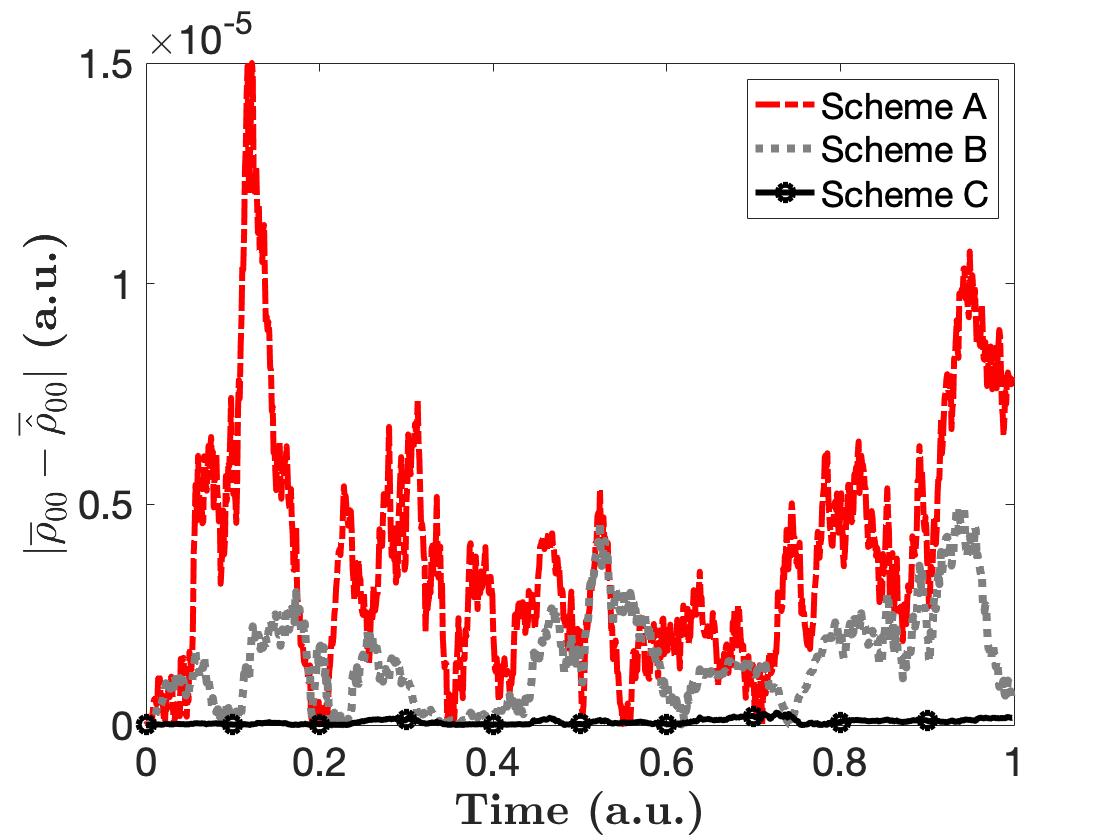}%
 \caption{The error in the $\rho_{00}$ entry of the density-matrix. Scheme I (red dashed); Scheme II (gray dot); Scheme III (black solid). }%
 \end{figure}

In Mora \cite{mora2005numerical}, an exponential Scheme called Euler-exponential for SSE \eqref{eq: sse}  is proposed,  which has weak 1.0 order convergence. It can be written as
\begin{equation}
\psi^{n+1} = \mathcal{P}\Big(\exp((-iH-\frac{1}{2}\hat{V}^*\hat{V})\Delta t) (\psi^n + \hat{V}\Delta W \psi^n)\Big),
\end{equation} 
where $\mathcal{P}$ is the projection to the unit ball to ensure the norm-preserving property. As a comparison, Fig 4 depicts the error from the Euler-Exponential method (the blue solid line on top), compared to our schemes. Our schemes yield significantly smaller error than the Euler-Exponential method.  
  \begin{figure}[H]
  \label{fig3}
  \centering
 \includegraphics[scale = 0.2]{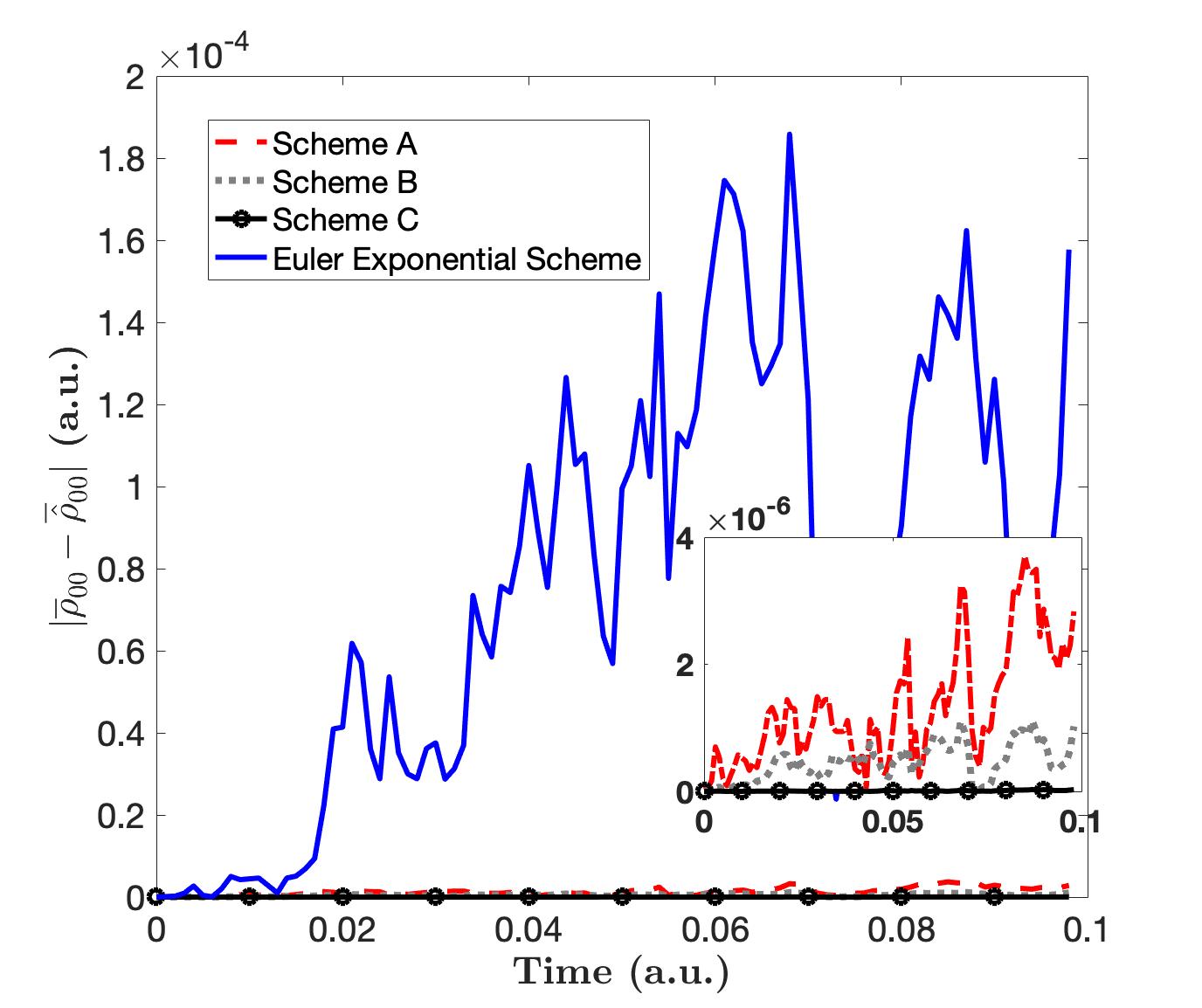}%
 \caption{A comparison to the Euler exponential scheme (blue solid) constructed by Mora \cite{mora2005numerical}. Scheme I (red dashed), II (gray dot) and III (black solid). }%
 \end{figure}

Now we consider the nonlinear case, and we pick $\tilde{g} = 1$ in \eqref{eq: SSEC}. As we have discussed in Section II B, we adopted the symmetric splitting scheme. The exact solution is computed by the Euler-Maruyama method with a much smaller time step $\delta t = 10^{-3} \Delta t$.  We approximate the expectation using 100 runs.  From FIG.4 and FIG.5, we make similar observations as in FIG.1 and FIG.2.

 \begin{figure}[H]
 \centering
 \label{fig3}
 \includegraphics[scale = 0.2]{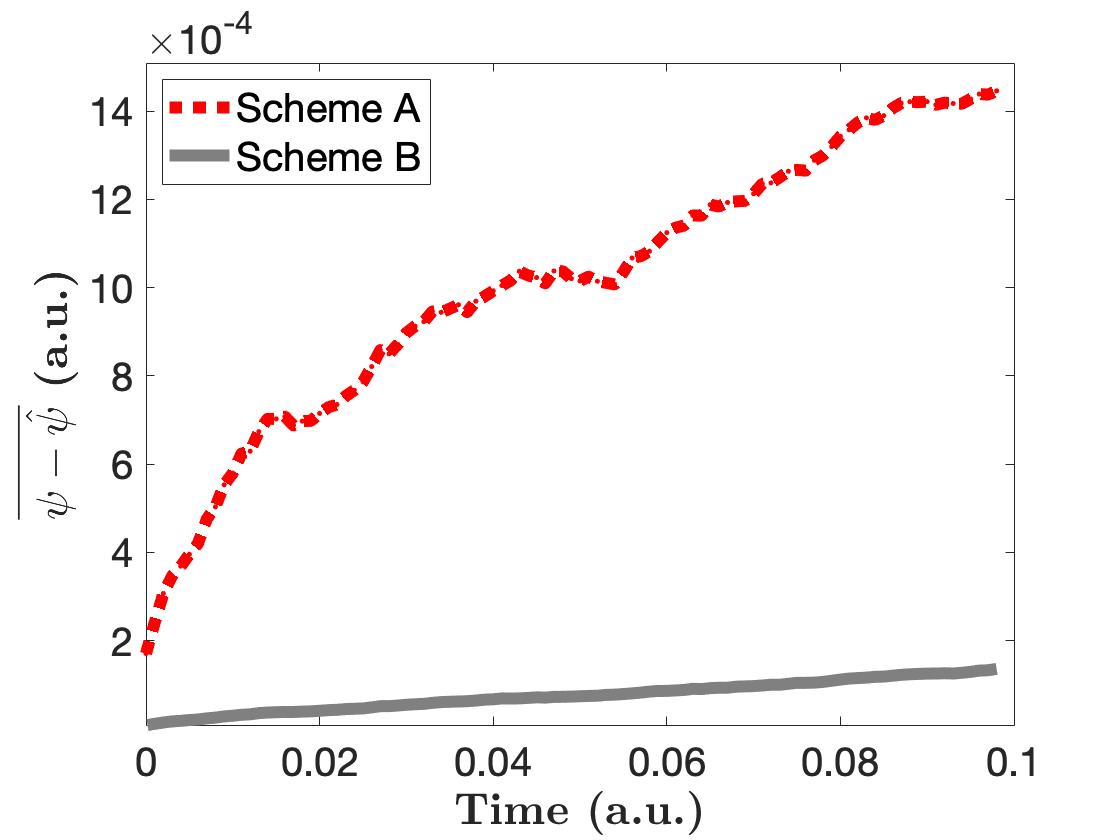}%
 \caption{A comparison of the strong order of Scheme I and II in the nonlinear case: $\tilde g = 1$. Scheme I (red dot); Scheme II (gray solid).   }%
 \end{figure}

 \begin{figure}[H]
 \centering
 \label{fig3}
 \includegraphics[scale = 0.2]{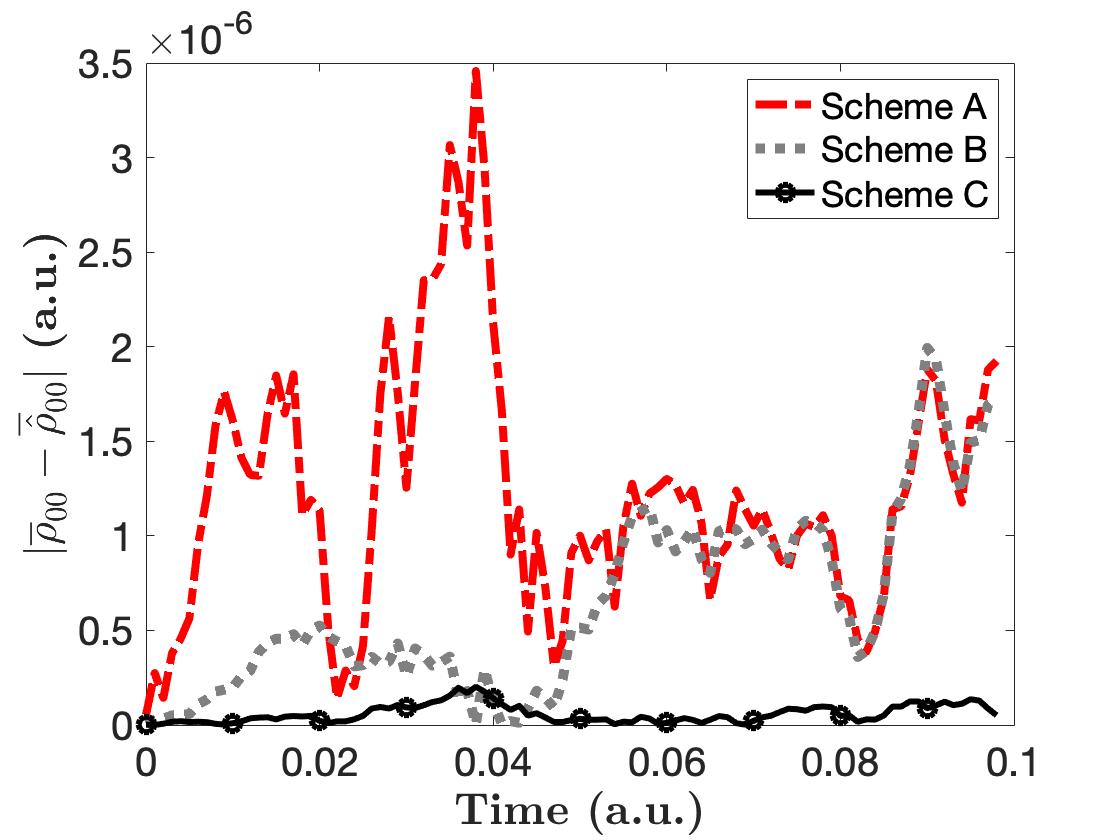}%
 \caption{The error of the entry $\rho_{11}$ of the density-matrix  in the nonlinear case: $\tilde g = 1$. Scheme I (red solid); Scheme II (gray solid)l; Scheme III (black solid).  }%
 \end{figure}


\medskip
One of the crucial properties of the SSE is the mass conservation. This implies that $\overline{\|\psi(t)\|^2}$ should remain constant. To test this property, we test the exponential integrator  on a long time period and compare it with the Euler-Maruyama method. We choose the following bath operator,
 \begin{equation}\label{eq: V2}
 \begin{aligned}
\hat{V}_2 \equiv \frac{1}{10} 
 \begin{pmatrix} 
            0 & 1 & 1 &1 & \cdots \\
            0 & 0 & 0 &0 & \cdots\\
            \vdots & \vdots & \vdots &\vdots & \vdots\\
            0& 0 & 0 &0 &0\\
            1 & 1 & \cdots & 1 & 0
 \end{pmatrix},
\end{aligned}\end{equation}

We pick time step $\Delta t = 10^{-4}$ , ensuring that the Euler-Maruyama method is stable under this setting.  And we involve the system for $10^5$ time steps to $T=10$. The ensemble average is approximated by averaging over 1000 runs, and doing a larger ensemble size did not result in noticeable changes. We depict the norm from the two methods in FIG. 6. The norm of Scheme II and III are very close to Scheme I so we omit them in the figure. We observe that the Euler-Maruyama method causes the norm to increase quite quickly, while the norm from  schemes I, II and III  seem to decrease, but it remains much closer to 1.

 \begin{figure}[H]
 \centering
 \label{fig3}
 \includegraphics[scale = 0.2]{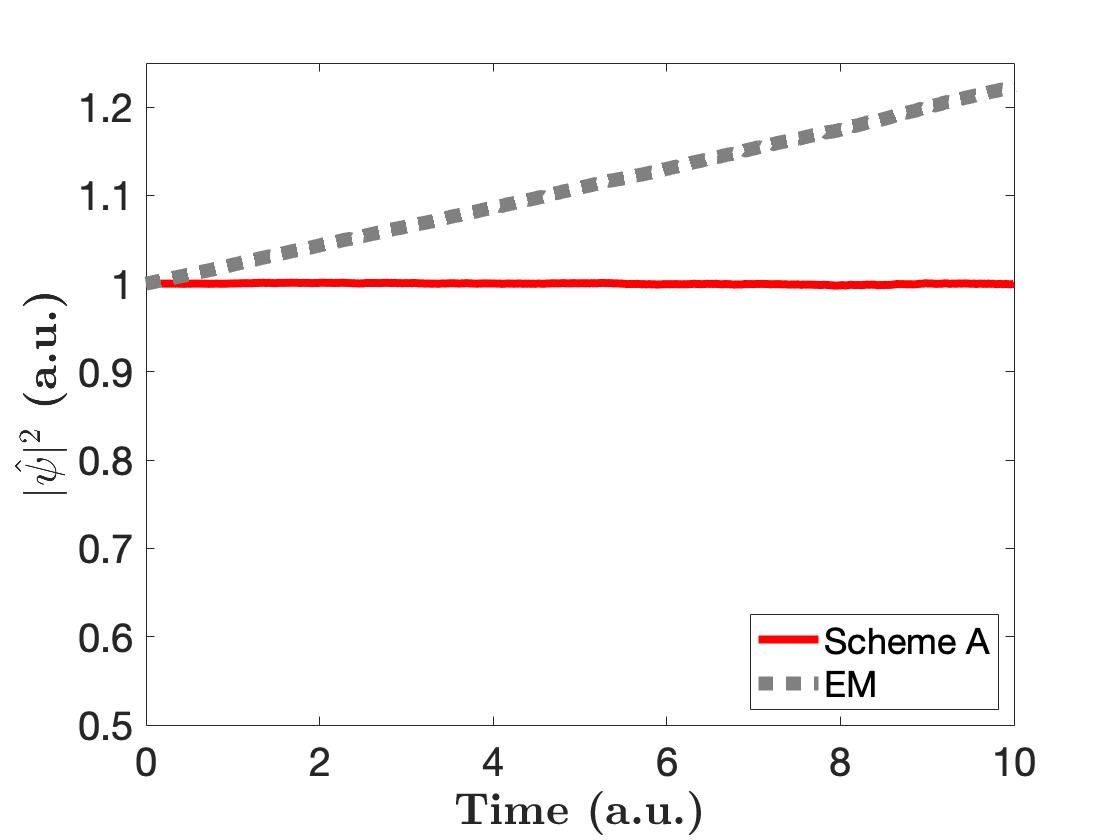}%
 \caption{The ensemble average of the norm of state vector (total mass). Scheme I (red solid); Euler-Maruyama (gray dash).  }%
 \end{figure}

\section{Summary and Discussion}

In this paper we proposed  exponential integrators for the  stochastic Schr\"{o}dinger equation based on Kunita's representation \cite{kunita1980representation}, which can be efficiently implemented by using the Krylov subspace method. Our schemes I and II have been verified to have order 1.0 and order 1.5, respectively, in the sense of strong convergence.  We also discussed their convergence in terms of  the density-matrix. This analysis also suggests that the accuracy can be improved by adding  two commutators in the Scheme II.  The nonlinear case is addressed by adopting an operator-splitting method \cite{watanabe2002efficient, misawa2000numerical, misawa2001lie}, where the linear and nonlinear parts are treated separately. 

In our numerical tests, we have found that our exponential schemes have better stability property, when compared to the Euler-Maruyama and Milstein's methods. They are also much better at preserving the square norm (mass) of the wave function.  Overall, these schemes are good alternatives  in the computation of the stochastic Schr\"{o}dinger equation. 
Meanwhile, unlike some of the methods for deterministic Schr\"{o}dinger equations,  these methods do not exactly preserve the norm.  It is still an open challenge to find a norm-preserving integrator.

\section{Appendix}

\subsection{The Proof of Theorem II.1}

\begin{proof}

We can actually use Theorem II.2 to prove this lemma.  $\hat{V}\equiv 0$ gives the deterministic case. 
Since the It\^{o} formula\cite{oksendal2003stochastic} is defined for real-valued equations, a simple idea is to separate the state vector into real and  imaginary parts, 
\begin{equation}\label{eq:realSE}\begin{aligned} 
   d   \begin{bmatrix} 
         u  \\
         v  \\
      \end{bmatrix} &=     
      \begin{bmatrix} 
            0 & H \\
            -H &   0\\
         \end{bmatrix}
         \begin{bmatrix} 
         u  \\
         v  \\
      \end{bmatrix} dt,
\end{aligned}\end{equation}
where $u$ and $v$ are respectively the real and imaginary part of $\psi(x,t)$.
Theorem II.2 gives the following representation to the solution of \eqref{eq:realSE} 

\begin{equation}\begin{aligned} 
      \begin{bmatrix} 
         u_t  \\
         v_t  \\
      \end{bmatrix} &= 
      \exp(D_t)
      \begin{bmatrix} 
         u  \\
         v  \\
      \end{bmatrix}  ,     \end{aligned}\end{equation}

where 
\begin{equation}\label{eq:Koopman}\begin{aligned} 
   D_t &= tHv\frac{\partial}{\partial u} -tHu\frac{\partial}{\partial v}.
\end{aligned}\end{equation}

By taking $f(u,v) = u+iv = \psi$, we have,
\begin{equation}\begin{aligned} 
   \psi(x,t) &=  \exp{D_t}\psi.
\end{aligned}\end{equation}

According to the chain rule \cite{watanabe2002efficient}, 
\begin{equation}\begin{aligned} 
   \frac{\partial}{\partial u} &= \frac{\partial \psi}{\partial u}\frac{\partial }{\partial \psi}  +  \frac{\partial \psi^*}{\partial u}\frac{\partial }{\partial \psi^*} = \frac{\partial }{\partial \psi}  + \frac{\partial }{\partial \psi^*}, \\
   \frac{\partial}{\partial v} &= \frac{\partial \psi}{\partial v}\frac{\partial }{\partial \psi}  +  \frac{\partial \psi^*}{\partial v}\frac{\partial }{\partial \psi^*} = i\frac{\partial }{\partial \psi}  -i \frac{\partial }{\partial \psi^*}, 
\end{aligned}\end{equation}
and a substitution into \eqref{eq:Koopman} , one gets
\begin{equation}\begin{aligned} 
   D_t &=  tHv( \frac{\partial }{\partial \psi}  + \frac{\partial }{\partial \psi^*} ) -tHu(i\frac{\partial }{\partial \psi}  -i \frac{\partial }{\partial \psi^*} ),\\
   &= -itH(u+iv)\frac{\partial }{\partial \psi}  + (-itH)(-u+iv)\frac{\partial }{\partial \psi^*} ,\\
   &=-itH(\psi\frac{\partial }{\partial \psi} - \psi^*\frac{\partial }{\partial \psi^*} ).
   \end{aligned}\end{equation}

For the second part, since $H$ is  linear  and  $\frac{\partial }{\partial \psi^*} \psi = 0$,
\begin{equation} 
   \exp(D_t)\psi =   \sum_{j= 0}^{\infty}\frac{1}{j!}[-itH(\psi\frac{\partial }{\partial \psi} - \psi^*\frac{\partial }{\partial \psi^*} )]^j \psi
    = \exp(-itH)\psi.
\end{equation}

\end{proof}

\subsection{Proof of Corollary II.3}

\begin{proof}

We first rewrite the equation as
\begin{equation}\label{eq:realSSE'}\begin{aligned} 
   d   \begin{bmatrix} 
         u  \\
         v  \\
      \end{bmatrix} &=     
      \begin{bmatrix} 
            -M & H \\
            -H &   -M\\
         \end{bmatrix}
         \begin{bmatrix} 
         u  \\
         v  \\
      \end{bmatrix} dt
      +    
       \begin{bmatrix} 
               \hat{V} & 0 \\
               0 & \hat{V} \\
            \end{bmatrix}
            \begin{bmatrix} 
         u  \\
         v  \\
      \end{bmatrix}\circ dW,
\end{aligned}\end{equation}
where
\begin{equation}\begin{aligned} 
   M &:=  \frac{1}{2}(\hat{V}^*+ \hat{V} )\hat{V} .
\end{aligned}\end{equation}

Theorem II.2 gives the following representation to the solution of \eqref{eq:realSSE'} 
\begin{equation}\label{eq:Koopman'}\begin{aligned} 
      \begin{bmatrix} 
         u_t  \\
         v_t  \\
      \end{bmatrix} &= 
      \exp{\tilde{D}_t}
      \begin{bmatrix} 
         u  \\
         v  \\
      \end{bmatrix},       \end{aligned}\end{equation}
where $\tilde{D}_t$ is expressed as a Magnus expansion. If we take the truncation of the first two terms, we have,
\begin{equation}\begin{aligned} 
   \tilde{D}_{ t}^{I} &=  
   \Bigg \{
    \begin{bmatrix} 
            -M & H \\
            -H &   -M\\
         \end{bmatrix}
          \begin{bmatrix} 
         u  \\
         v  \\
      \end{bmatrix} J_{(0)}
      + 
      \begin{bmatrix} 
               \hat{V} & 0 \\
               0 & \hat{V} \\
            \end{bmatrix}
            \begin{bmatrix} 
         u  \\
         v  \\
      \end{bmatrix} J_{(1)}
       \Bigg \} \cdot \nabla
\end{aligned}\end{equation} 

To be concise, we let 
\begin{equation}\begin{aligned} 
   A &:=  -Mt +  \hat{V}W_t,\\
   B &:=  Ht.
\end{aligned}\end{equation}
As a result, the truncated operator can be written in a more compact form,
\begin{equation}\begin{aligned} 
\tilde{D}_{ t}^{I} &=  
 \begin{bmatrix} 
            A & B \\
            -B &  A\\
         \end{bmatrix}
          \begin{bmatrix} 
         u  \\
         v  \\
      \end{bmatrix}  \cdot \nabla
      =   (Au + Bv) \frac{\partial}{\partial u} + (-Bu+Av)\frac{\partial}{\partial v}.
\end{aligned}\end{equation}

Following the same idea  in Themreo II.1, we get
\begin{equation}\begin{aligned} 
  \tilde{D}_{ t}^{I}  &=  (Au + Bv)(\frac{\partial }{\partial \psi}  + \frac{\partial }{\partial \psi^*}) + (-Bu+Av)(i\frac{\partial }{\partial \psi}  -i \frac{\partial }{\partial \psi^*} )\\
  &=  (A-iB)\psi\frac{\partial }{\partial \psi} + (A+iB)\psi^*\frac{\partial }{\partial \psi^*}.
\end{aligned}\end{equation}

Since $\frac{\partial }{\partial \psi^*} \psi = 0$, we have
\begin{equation} 
   \psi^I(x,t) =  \exp(\tilde{D}_{ t}^{I} )\psi
               = \exp\{(-iH- \frac{1}{2}(\hat{V}^*+ \hat{V} )\hat{V}) t + \hat{V}W_t\}\psi.
\end{equation}

Similarly, if we include the 3rd term of the expansion of $\tilde{D_t}$ in \eqref{eq:Koopman'}, \ie,
\begin{equation}\begin{aligned} 
   A &:=  -MJ_{(0)} +  \hat{V} J_{(1)}+ \frac{1}{2}\{J_{(0,1)}-J_{(1,0)}\}[\hat{V},M],\\
   B &:=  H J_{(0)} +  \frac{1}{2}\{J_{(0,1)}-J_{(1,0)}\}[\hat{V},H],
\end{aligned}\end{equation}

we have
\begin{equation}\begin{aligned} 
   \psi^{II}(x,t) &=  \exp(\tilde{D}_{ t}^{I} )\psi,\\
               &= \exp\{(-iH- \frac{1}{2}(\hat{V}^*+ \hat{V} )\hat{V} )t + \hat{V}W_t\\
           &+  \frac{1}{2}(J_{(0,1)}-J_{(1,0)}) (\frac{1}{2}[\hat{V}^*, \hat{V}] \hat{V}+ i[H,\hat{V}])\}\psi.
\end{aligned}\end{equation}

This verifies the claim. 

\end{proof}

\subsection{The Proof of Theorem II.5}

\begin{proof}

We can write the equation for the density-matrix
\begin{eqnarray} 
  \partial_t \hat{ \rho}= \exp(A\Delta t + B\Delta W + C\Delta U) \hat{\rho}_{t_0} \exp(A^*\Delta t + B^*\Delta W + C^*\Delta U),
\end{eqnarray}
from the exponential integrators. We can expand the exponential, and using
\begin{eqnarray} 
 \overline{\Delta W \Delta W} = && \Delta t ,
\end{eqnarray}
we have 
\begin{eqnarray} 
\partial_t\hat{\rho}(0) = &&A\rho(0) + \rho(0)A^* + B\rho(0)B^* + \frac{1}{2}BB\rho(0) + \frac{1}{2}\rho(0)B^*B^*
:= \mathcal{L}\rho(0)
\end{eqnarray}

This agrees with the first derivative of the density-matrix computed from the Lindblad equation.  Since the $\Delta U$ term does not contribute to the order $\mathcal{O}(\Delta t)$ term, we have $\partial_{t}\rho(0) = \partial_{t} \hat{\rho}_{I}(0)= \partial_{t} \hat{\rho}_{II}(0)$. 

From the Lindblad equation, we can also compute the second derivative of the density-matrix:
\begin{eqnarray} 
\partial_{tt}\rho(0) = \mathcal{L}\mathcal{L}\rho(0).
\end{eqnarray}

We can expand it with some lengthy calculation:
\begin{equation} 
\begin{aligned}
\partial_{tt}\rho &=  AA\rho + \rho A^*A^* + 2A\rho A^* \\
&+ AB\rho B^* + BA\rho B^* + A\rho B^*B^* +B\rho B^*A^* + B\rho A^*B^* +  B\rho B^*A^* \\
& + \frac{1}{2}ABB\rho + \frac{1}{2}BBA\rho + \frac{1}{2}\rho A^*B^*B^* + \frac{1}{2}\rho B^*B^*A^*\\
& + \frac{1}{4}BBBB\rho + BBB\rho B^* + \frac{3}{2}BB\rho B^*B^* + B\rho B^*B^*B^* + \frac{1}{4}\rho B^*B^*B^*B^*.
\end{aligned}
\end{equation}

This is the same as $\partial_{tt}\hat{\rho}_{III}(0)$. Thus we have 
$\partial_{tt}\rho(0) = \partial_{tt} \hat{\rho}_{III}(0)$.

\end{proof}


\bibliography{scholar,openqm}

\providecommand{\noopsort}[1]{}\providecommand{\singleletter}[1]{#1}
\begin{thebibliography}{31}%
\makeatletter
\providecommand \@ifxundefined [1]{%
 \@ifx{#1\undefined}
}%
\providecommand \@ifnum [1]{%
 \ifnum #1\expandafter \@firstoftwo
 \else \expandafter \@secondoftwo
 \fi
}%
\providecommand \@ifx [1]{%
 \ifx #1\expandafter \@firstoftwo
 \else \expandafter \@secondoftwo
 \fi
}%
\providecommand \natexlab [1]{#1}%
\providecommand \enquote  [1]{``#1''}%
\providecommand \bibnamefont  [1]{#1}%
\providecommand \bibfnamefont [1]{#1}%
\providecommand \citenamefont [1]{#1}%
\providecommand \href@noop [0]{\@secondoftwo}%
\providecommand \href [0]{\begingroup \@sanitize@url \@href}%
\providecommand \@href[1]{\@@startlink{#1}\@@href}%
\providecommand \@@href[1]{\endgroup#1\@@endlink}%
\providecommand \@sanitize@url [0]{\catcode `\\12\catcode `\$12\catcode
  `\&12\catcode `\#12\catcode `\^12\catcode `\_12\catcode `\%12\relax}%
\providecommand \@@startlink[1]{}%
\providecommand \@@endlink[0]{}%
\providecommand \url  [0]{\begingroup\@sanitize@url \@url }%
\providecommand \@url [1]{\endgroup\@href {#1}{\urlprefix }}%
\providecommand \urlprefix  [0]{URL }%
\providecommand \Eprint [0]{\href }%
\providecommand \doibase [0]{http://dx.doi.org/}%
\providecommand \selectlanguage [0]{\@gobble}%
\providecommand \bibinfo  [0]{\@secondoftwo}%
\providecommand \bibfield  [0]{\@secondoftwo}%
\providecommand \translation [1]{[#1]}%
\providecommand \BibitemOpen [0]{}%
\providecommand \bibitemStop [0]{}%
\providecommand \bibitemNoStop [0]{.\EOS\space}%
\providecommand \EOS [0]{\spacefactor3000\relax}%
\providecommand \BibitemShut  [1]{\csname bibitem#1\endcsname}%
\let\auto@bib@innerbib\@empty
\bibitem [{\citenamefont {{Breuer, Heinz-Peter and Petruccione, Francesco and
  others}}(2002)}]{breuer2002theory}%
  \BibitemOpen
  \bibfield  {author} {\bibinfo {author} {\bibnamefont {{Breuer, Heinz-Peter
  and Petruccione, Francesco and others}}},\ }\href@noop {} {\emph {\bibinfo
  {title} {The theory of open quantum systems}}}\ (\bibinfo  {publisher}
  {Oxford University Press on Demand},\ \bibinfo {year} {2002})\BibitemShut
  {NoStop}%
\bibitem [{\citenamefont {Carmichael}(2009)}]{carmichael2009open}%
  \BibitemOpen
  \bibfield  {author} {\bibinfo {author} {\bibfnamefont {H.}~\bibnamefont
  {Carmichael}},\ }\href@noop {} {\emph {\bibinfo {title} {An open systems
  approach to quantum optics: lectures presented at the Universit{\'e} Libre de
  Bruxelles, October 28 to November 4, 1991}}},\ Vol.~\bibinfo {volume} {18}\
  (\bibinfo  {publisher} {Springer Science \& Business Media},\ \bibinfo {year}
  {2009})\BibitemShut {NoStop}%
\bibitem [{\citenamefont {Weiss}(2012)}]{weiss2012quantum}%
  \BibitemOpen
  \bibfield  {author} {\bibinfo {author} {\bibfnamefont {U.}~\bibnamefont
  {Weiss}},\ }\href@noop {} {\emph {\bibinfo {title} {Quantum dissipative
  systems}}},\ Vol.~\bibinfo {volume} {13}\ (\bibinfo  {publisher} {World
  scientific},\ \bibinfo {year} {2012})\BibitemShut {NoStop}%
\bibitem [{\citenamefont {Gaspard}\ and\ \citenamefont
  {Nagaoka}(1999)}]{gaspard1999non}%
  \BibitemOpen
  \bibfield  {author} {\bibinfo {author} {\bibfnamefont {P.}~\bibnamefont
  {Gaspard}}\ and\ \bibinfo {author} {\bibfnamefont {M.}~\bibnamefont
  {Nagaoka}},\ }\bibfield  {title} {\enquote {\bibinfo {title} {Non-markovian
  stochastic {S}chr{\"o}dinger equation},}\ }\href@noop {} {\bibfield
  {journal} {\bibinfo  {journal} {The Journal of chemical physics}\ }\textbf
  {\bibinfo {volume} {111}},\ \bibinfo {pages} {5676--5690} (\bibinfo {year}
  {1999})}\BibitemShut {NoStop}%
\bibitem [{\citenamefont {Lindblad}(1976)}]{lindblad1976generators}%
  \BibitemOpen
  \bibfield  {author} {\bibinfo {author} {\bibfnamefont {G.}~\bibnamefont
  {Lindblad}},\ }\bibfield  {title} {\enquote {\bibinfo {title} {On the
  generators of quantum dynamical semigroups},}\ }\href@noop {} {\bibfield
  {journal} {\bibinfo  {journal} {Communications in Mathematical Physics}\
  }\textbf {\bibinfo {volume} {48}},\ \bibinfo {pages} {119--130} (\bibinfo
  {year} {1976})}\BibitemShut {NoStop}%
\bibitem [{\citenamefont {Di~Ventra}\ and\ \citenamefont
  {Todorov}(2004)}]{di2004transport}%
  \BibitemOpen
  \bibfield  {author} {\bibinfo {author} {\bibfnamefont {M.}~\bibnamefont
  {Di~Ventra}}\ and\ \bibinfo {author} {\bibfnamefont {T.~N.}\ \bibnamefont
  {Todorov}},\ }\bibfield  {title} {\enquote {\bibinfo {title} {Transport in
  nanoscale systems: the microcanonical versus grand-canonical picture},}\
  }\href@noop {} {\bibfield  {journal} {\bibinfo  {journal} {Journal of
  Physics: Condensed Matter}\ }\textbf {\bibinfo {volume} {16}},\ \bibinfo
  {pages} {8025} (\bibinfo {year} {2004})}\BibitemShut {NoStop}%
\bibitem [{\citenamefont {Di~Ventra}\ and\ \citenamefont
  {D’Agosta}(2007)}]{di2007stochastic}%
  \BibitemOpen
  \bibfield  {author} {\bibinfo {author} {\bibfnamefont {M.}~\bibnamefont
  {Di~Ventra}}\ and\ \bibinfo {author} {\bibfnamefont {R.}~\bibnamefont
  {D’Agosta}},\ }\bibfield  {title} {\enquote {\bibinfo {title} {Stochastic
  time-dependent current-density-functional theory},}\ }\href@noop {}
  {\bibfield  {journal} {\bibinfo  {journal} {Physical review letters}\
  }\textbf {\bibinfo {volume} {98}},\ \bibinfo {pages} {226403} (\bibinfo
  {year} {2007})}\BibitemShut {NoStop}%
\bibitem [{\citenamefont {D’Agosta}\ and\ \citenamefont
  {Di~Ventra}(2008)}]{d2008stochastic}%
  \BibitemOpen
  \bibfield  {author} {\bibinfo {author} {\bibfnamefont {R.}~\bibnamefont
  {D’Agosta}}\ and\ \bibinfo {author} {\bibfnamefont {M.}~\bibnamefont
  {Di~Ventra}},\ }\bibfield  {title} {\enquote {\bibinfo {title} {Stochastic
  time-dependent current-density-functional theory: A functional theory of open
  quantum systems},}\ }\href@noop {} {\bibfield  {journal} {\bibinfo  {journal}
  {Physical Review B}\ }\textbf {\bibinfo {volume} {78}},\ \bibinfo {pages}
  {165105} (\bibinfo {year} {2008})}\BibitemShut {NoStop}%
\bibitem [{\citenamefont {Biele}\ and\ \citenamefont
  {D’Agosta}(2012)}]{biele2012stochastic}%
  \BibitemOpen
  \bibfield  {author} {\bibinfo {author} {\bibfnamefont {R.}~\bibnamefont
  {Biele}}\ and\ \bibinfo {author} {\bibfnamefont {R.}~\bibnamefont
  {D’Agosta}},\ }\bibfield  {title} {\enquote {\bibinfo {title} {A stochastic
  approach to open quantum systems},}\ }\href@noop {} {\bibfield  {journal}
  {\bibinfo  {journal} {Journal of Physics: Condensed Matter}\ }\textbf
  {\bibinfo {volume} {24}},\ \bibinfo {pages} {273201} (\bibinfo {year}
  {2012})}\BibitemShut {NoStop}%
\bibitem [{\citenamefont {Castro}, \citenamefont {Marques},\ and\ \citenamefont
  {Rubio}(2004)}]{castro2004propagators}%
  \BibitemOpen
  \bibfield  {author} {\bibinfo {author} {\bibfnamefont {A.}~\bibnamefont
  {Castro}}, \bibinfo {author} {\bibfnamefont {M.~A.}\ \bibnamefont {Marques}},
  \ and\ \bibinfo {author} {\bibfnamefont {A.}~\bibnamefont {Rubio}},\
  }\bibfield  {title} {\enquote {\bibinfo {title} {Propagators for the
  time-dependent {Kohn--Sham} equations},}\ }\href@noop {} {\bibfield
  {journal} {\bibinfo  {journal} {The Journal of chemical physics}\ }\textbf
  {\bibinfo {volume} {121}},\ \bibinfo {pages} {3425--3433} (\bibinfo {year}
  {2004})}\BibitemShut {NoStop}%
\bibitem [{\citenamefont {Saad}(1992)}]{saad1992analysis}%
  \BibitemOpen
  \bibfield  {author} {\bibinfo {author} {\bibfnamefont {Y.}~\bibnamefont
  {Saad}},\ }\bibfield  {title} {\enquote {\bibinfo {title} {Analysis of some
  {Krylov} subspace approximations to the matrix exponential operator},}\
  }\href@noop {} {\bibfield  {journal} {\bibinfo  {journal} {SIAM Journal on
  Numerical Analysis}\ }\textbf {\bibinfo {volume} {29}},\ \bibinfo {pages}
  {209--228} (\bibinfo {year} {1992})}\BibitemShut {NoStop}%
\bibitem [{\citenamefont {Gene}\ and\ \citenamefont
  {Charles}(1996)}]{gene1996matrix}%
  \BibitemOpen
  \bibfield  {author} {\bibinfo {author} {\bibfnamefont {H.~G.}\ \bibnamefont
  {Gene}}\ and\ \bibinfo {author} {\bibfnamefont {F.}~\bibnamefont {Charles}},\
  }\bibfield  {title} {\enquote {\bibinfo {title} {Matrix computations},}\
  }\href@noop {} {\bibfield  {journal} {\bibinfo  {journal} {Johns Hopkins
  Universtiy Press, 3rd edtion}\ } (\bibinfo {year} {1996})}\BibitemShut
  {NoStop}%
\bibitem [{\citenamefont {Burrage}\ \emph {et~al.}(2006)\citenamefont
  {Burrage}, \citenamefont {Burrage}, \citenamefont {Higham}, \citenamefont
  {Kloeden},\ and\ \citenamefont {Platen}}]{burrage2006comment}%
  \BibitemOpen
  \bibfield  {author} {\bibinfo {author} {\bibfnamefont {K.}~\bibnamefont
  {Burrage}}, \bibinfo {author} {\bibfnamefont {P.}~\bibnamefont {Burrage}},
  \bibinfo {author} {\bibfnamefont {D.~J.}\ \bibnamefont {Higham}}, \bibinfo
  {author} {\bibfnamefont {P.~E.}\ \bibnamefont {Kloeden}}, \ and\ \bibinfo
  {author} {\bibfnamefont {E.}~\bibnamefont {Platen}},\ }\bibfield  {title}
  {\enquote {\bibinfo {title} {Comment on “numerical methods for stochastic
  differential equations”},}\ }\href@noop {} {\bibfield  {journal} {\bibinfo
  {journal} {Physical Review E}\ }\textbf {\bibinfo {volume} {74}},\ \bibinfo
  {pages} {068701} (\bibinfo {year} {2006})}\BibitemShut {NoStop}%
\bibitem [{\citenamefont {Kloeden}\ and\ \citenamefont
  {Platen}(2013)}]{kloeden2013numerical}%
  \BibitemOpen
  \bibfield  {author} {\bibinfo {author} {\bibfnamefont {P.~E.}\ \bibnamefont
  {Kloeden}}\ and\ \bibinfo {author} {\bibfnamefont {E.}~\bibnamefont
  {Platen}},\ }\href@noop {} {\emph {\bibinfo {title} {Numerical solution of
  stochastic differential equations}}},\ Vol.~\bibinfo {volume} {23}\ (\bibinfo
   {publisher} {Springer Science \& Business Media},\ \bibinfo {year}
  {2013})\BibitemShut {NoStop}%
\bibitem [{\citenamefont {Hochbruck}\ and\ \citenamefont
  {Lubich}(1997)}]{hochbruck1997krylov}%
  \BibitemOpen
  \bibfield  {author} {\bibinfo {author} {\bibfnamefont {M.}~\bibnamefont
  {Hochbruck}}\ and\ \bibinfo {author} {\bibfnamefont {C.}~\bibnamefont
  {Lubich}},\ }\bibfield  {title} {\enquote {\bibinfo {title} {On {Krylov}
  subspace approximations to the matrix exponential operator},}\ }\href@noop {}
  {\bibfield  {journal} {\bibinfo  {journal} {SIAM Journal on Numerical
  Analysis}\ }\textbf {\bibinfo {volume} {34}},\ \bibinfo {pages} {1911--1925}
  (\bibinfo {year} {1997})}\BibitemShut {NoStop}%
\bibitem [{\citenamefont {Kunita}(1980)}]{kunita1980representation}%
  \BibitemOpen
  \bibfield  {author} {\bibinfo {author} {\bibfnamefont {H.}~\bibnamefont
  {Kunita}},\ }\bibfield  {title} {\enquote {\bibinfo {title} {On the
  representation of solutions of stochastic differential equations},}\ }in\
  \href@noop {} {\emph {\bibinfo {booktitle} {S{\'e}minaire de Probabilit{\'e}s
  XIV 1978/79}}}\ (\bibinfo  {publisher} {Springer},\ \bibinfo {year} {1980})\
  pp.\ \bibinfo {pages} {282--304}\BibitemShut {NoStop}%
\bibitem [{\citenamefont {Saad}(2003)}]{saad2003iterative}%
  \BibitemOpen
  \bibfield  {author} {\bibinfo {author} {\bibfnamefont {Y.}~\bibnamefont
  {Saad}},\ }\href@noop {} {\emph {\bibinfo {title} {Iterative methods for
  sparse linear systems}}},\ Vol.~\bibinfo {volume} {82}\ (\bibinfo
  {publisher} {siam},\ \bibinfo {year} {2003})\BibitemShut {NoStop}%
\bibitem [{\citenamefont {{\O}ksendal}(2003)}]{oksendal2003stochastic}%
  \BibitemOpen
  \bibfield  {author} {\bibinfo {author} {\bibfnamefont {B.}~\bibnamefont
  {{\O}ksendal}},\ }\bibfield  {title} {\enquote {\bibinfo {title} {Stochastic
  differential equations},}\ }in\ \href@noop {} {\emph {\bibinfo {booktitle}
  {Stochastic differential equations}}}\ (\bibinfo  {publisher} {Springer},\
  \bibinfo {year} {2003})\ pp.\ \bibinfo {pages} {65--84}\BibitemShut {NoStop}%
\bibitem [{\citenamefont {Misawa}(2001)}]{misawa2001lie}%
  \BibitemOpen
  \bibfield  {author} {\bibinfo {author} {\bibfnamefont {T.}~\bibnamefont
  {Misawa}},\ }\bibfield  {title} {\enquote {\bibinfo {title} {A {Lie}
  algebraic approach to numerical integration of stochastic differential
  equations},}\ }\href@noop {} {\bibfield  {journal} {\bibinfo  {journal} {SIAM
  Journal on Scientific Computing}\ }\textbf {\bibinfo {volume} {23}},\
  \bibinfo {pages} {866--890} (\bibinfo {year} {2001})}\BibitemShut {NoStop}%
\bibitem [{\citenamefont {Watanabe}\ and\ \citenamefont
  {Tsukada}(2002)}]{watanabe2002efficient}%
  \BibitemOpen
  \bibfield  {author} {\bibinfo {author} {\bibfnamefont {N.}~\bibnamefont
  {Watanabe}}\ and\ \bibinfo {author} {\bibfnamefont {M.}~\bibnamefont
  {Tsukada}},\ }\bibfield  {title} {\enquote {\bibinfo {title} {Efficient
  method for simulating quantum electron dynamics under the time-dependent
  {Kohn-Sham} equation},}\ }\href@noop {} {\bibfield  {journal} {\bibinfo
  {journal} {Physical Review E}\ }\textbf {\bibinfo {volume} {65}},\ \bibinfo
  {pages} {036705} (\bibinfo {year} {2002})}\BibitemShut {NoStop}%
\bibitem [{\citenamefont {Koopman}(1931)}]{koopman1931hamiltonian}%
  \BibitemOpen
  \bibfield  {author} {\bibinfo {author} {\bibfnamefont {B.~O.}\ \bibnamefont
  {Koopman}},\ }\bibfield  {title} {\enquote {\bibinfo {title} {Hamiltonian
  systems and transformation in {Hilbert} space},}\ }\href@noop {} {\bibfield
  {journal} {\bibinfo  {journal} {Proceedings of the National Academy of
  Sciences of the United States of America}\ }\textbf {\bibinfo {volume}
  {17}},\ \bibinfo {pages} {315} (\bibinfo {year} {1931})}\BibitemShut
  {NoStop}%
\bibitem [{\citenamefont {Brunton}\ and\ \citenamefont
  {Kutz}(2019)}]{brunton2019data}%
  \BibitemOpen
  \bibfield  {author} {\bibinfo {author} {\bibfnamefont {S.~L.}\ \bibnamefont
  {Brunton}}\ and\ \bibinfo {author} {\bibfnamefont {J.~N.}\ \bibnamefont
  {Kutz}},\ }\href@noop {} {\emph {\bibinfo {title} {Data-driven science and
  engineering: Machine learning, dynamical systems, and control}}}\ (\bibinfo
  {publisher} {Cambridge University Press},\ \bibinfo {year}
  {2019})\BibitemShut {NoStop}%
\bibitem [{\citenamefont {Mauroy}\ and\ \citenamefont
  {Goncalves}(2016)}]{mauroy2016linear}%
  \BibitemOpen
  \bibfield  {author} {\bibinfo {author} {\bibfnamefont {A.}~\bibnamefont
  {Mauroy}}\ and\ \bibinfo {author} {\bibfnamefont {J.}~\bibnamefont
  {Goncalves}},\ }\bibfield  {title} {\enquote {\bibinfo {title} {Linear
  identification of nonlinear systems: A lifting technique based on the
  {Koopman} operator},}\ }in\ \href@noop {} {\emph {\bibinfo {booktitle} {2016
  IEEE 55th Conference on Decision and Control (CDC)}}}\ (\bibinfo
  {organization} {IEEE},\ \bibinfo {year} {2016})\ pp.\ \bibinfo {pages}
  {6500--6505}\BibitemShut {NoStop}%
\bibitem [{\citenamefont {Misawa}(2000)}]{misawa2000numerical}%
  \BibitemOpen
  \bibfield  {author} {\bibinfo {author} {\bibfnamefont {T.}~\bibnamefont
  {Misawa}},\ }\bibfield  {title} {\enquote {\bibinfo {title} {Numerical
  integration of stochastic differential equations by composition methods},}\
  }\href@noop {} {\bibfield  {journal} {\bibinfo  {journal}
  {Surikaisekikenkyusho Kokyuroku}\ }\textbf {\bibinfo {volume} {1180}},\
  \bibinfo {pages} {166--190} (\bibinfo {year} {2000})}\BibitemShut {NoStop}%
\bibitem [{\citenamefont {Telatovich}\ and\ \citenamefont
  {Li}(2017)}]{telatovich2017strong}%
  \BibitemOpen
  \bibfield  {author} {\bibinfo {author} {\bibfnamefont {A.}~\bibnamefont
  {Telatovich}}\ and\ \bibinfo {author} {\bibfnamefont {X.}~\bibnamefont
  {Li}},\ }\bibfield  {title} {\enquote {\bibinfo {title} {The strong
  convergence of operator-splitting methods for the {Langevin} dynamics
  model},}\ }\href@noop {} {\bibfield  {journal} {\bibinfo  {journal} {arXiv
  preprint arXiv:1706.04237}\ } (\bibinfo {year} {2017})}\BibitemShut {NoStop}%
\bibitem [{\citenamefont {Marques}\ \emph {et~al.}(2006)\citenamefont
  {Marques}, \citenamefont {Ullrich}, \citenamefont {Nogueira}, \citenamefont
  {Rubio}, \citenamefont {Burke},\ and\ \citenamefont
  {Gross}}]{marques2006time}%
  \BibitemOpen
  \bibfield  {author} {\bibinfo {author} {\bibfnamefont {M.~A.}\ \bibnamefont
  {Marques}}, \bibinfo {author} {\bibfnamefont {C.~A.}\ \bibnamefont
  {Ullrich}}, \bibinfo {author} {\bibfnamefont {F.}~\bibnamefont {Nogueira}},
  \bibinfo {author} {\bibfnamefont {A.}~\bibnamefont {Rubio}}, \bibinfo
  {author} {\bibfnamefont {K.}~\bibnamefont {Burke}}, \ and\ \bibinfo {author}
  {\bibfnamefont {E.~K.}\ \bibnamefont {Gross}},\ }\href@noop {} {\emph
  {\bibinfo {title} {{Time-dependent density functional theory}}}},\ Vol.\
  \bibinfo {volume} {706}\ (\bibinfo  {publisher} {Springer Science \& Business
  Media},\ \bibinfo {year} {2006})\BibitemShut {NoStop}%
\bibitem [{\citenamefont {Runge}\ and\ \citenamefont
  {Gross}(1984)}]{runge1984density}%
  \BibitemOpen
  \bibfield  {author} {\bibinfo {author} {\bibfnamefont {E.}~\bibnamefont
  {Runge}}\ and\ \bibinfo {author} {\bibfnamefont {E.~K.}\ \bibnamefont
  {Gross}},\ }\bibfield  {title} {\enquote {\bibinfo {title}
  {Density-functional theory for time-dependent systems},}\ }\href@noop {}
  {\bibfield  {journal} {\bibinfo  {journal} {Physical Review Letters}\
  }\textbf {\bibinfo {volume} {52}},\ \bibinfo {pages} {997} (\bibinfo {year}
  {1984})}\BibitemShut {NoStop}%
\bibitem [{\citenamefont {Yoshida}(1990)}]{yoshida1990construction}%
  \BibitemOpen
  \bibfield  {author} {\bibinfo {author} {\bibfnamefont {H.}~\bibnamefont
  {Yoshida}},\ }\bibfield  {title} {\enquote {\bibinfo {title} {Construction of
  higher order symplectic integrators},}\ }\href@noop {} {\bibfield  {journal}
  {\bibinfo  {journal} {Physics letters A}\ }\textbf {\bibinfo {volume}
  {150}},\ \bibinfo {pages} {262--268} (\bibinfo {year} {1990})}\BibitemShut
  {NoStop}%
\bibitem [{\citenamefont {Lord}, \citenamefont {Malham},\ and\ \citenamefont
  {Wiese}(2008)}]{lord2008efficient}%
  \BibitemOpen
  \bibfield  {author} {\bibinfo {author} {\bibfnamefont {G.}~\bibnamefont
  {Lord}}, \bibinfo {author} {\bibfnamefont {S.~J.}\ \bibnamefont {Malham}}, \
  and\ \bibinfo {author} {\bibfnamefont {A.}~\bibnamefont {Wiese}},\ }\bibfield
   {title} {\enquote {\bibinfo {title} {Efficient strong integrators for linear
  stochastic systems},}\ }\href@noop {} {\bibfield  {journal} {\bibinfo
  {journal} {SIAM Journal on Numerical Analysis}\ }\textbf {\bibinfo {volume}
  {46}},\ \bibinfo {pages} {2892--2919} (\bibinfo {year} {2008})}\BibitemShut
  {NoStop}%
\bibitem [{\citenamefont {Van~Kampen}(1992)}]{van1992stochastic}%
  \BibitemOpen
  \bibfield  {author} {\bibinfo {author} {\bibfnamefont {N.~G.}\ \bibnamefont
  {Van~Kampen}},\ }\href@noop {} {\emph {\bibinfo {title} {Stochastic processes
  in physics and chemistry}}},\ Vol.~\bibinfo {volume} {1}\ (\bibinfo
  {publisher} {Elsevier},\ \bibinfo {year} {1992})\BibitemShut {NoStop}%
\bibitem [{\citenamefont {Mora}(2005)}]{mora2005numerical}%
  \BibitemOpen
  \bibfield  {author} {\bibinfo {author} {\bibfnamefont {C.~M.}\ \bibnamefont
  {Mora}},\ }\bibfield  {title} {\enquote {\bibinfo {title} {Numerical solution
  of conservative finite-dimensional stochastic {Schr{\"o}dinger} equations},}\
  }\href@noop {} {\bibfield  {journal} {\bibinfo  {journal} {The Annals of
  Applied Probability}\ }\textbf {\bibinfo {volume} {15}},\ \bibinfo {pages}
  {2144--2171} (\bibinfo {year} {2005})}\BibitemShut {NoStop}%
\end{thebibliography}%
\end{document}